\documentstyle[11pt,newpasp,twoside,epsf]{article}
\def\edcomment#1{\iffalse\marginpar{\raggedright\sl#1\/}\else\relax\fi}
\reversemarginpar

\begin{document}
\title{Numerical Simulations of the pulsed Jet of MWC 560}
\author{Matthias Stute \& Max Camenzind}
\affil{Landessternwarte Heidelberg, K\"onigstuhl, D-69117 Heidelberg, Germany}
\author{Hans Martin Schmid}
\affil{Institute of Astronomy, ETH Zentrum, CH-8092 Zurich, Switzerland}

\begin{abstract}
    MWC 560 (= V694 Mon) is the only known Symbiotic Star system in 
    which the jet axis is practically parallel to the line of sight. Therefore 
    this system is predestinated to study the dynamical evolution and the 
    propagation of stellar jets. Spectroscopic monitoring done by Schmid et 
    al. (2001) showed that the outflow is seen as absorption features in the 
    continuum of the accretion disk and the accreting white dwarf, the 
    emission line spectrum of the accretion disk and the spectrum of the red 
    giant. We present the first numerical simulations of the jet of this 
    particular object using the {\em NIRVANA} code (Ziegler \& Yorke 1997) in 
    order to reproduce the velocity structures seen in the observational data. 
    This code solves the equations of hydrodynamics and was modified to
    calculate radiative losses due to non-equilibrium cooling by line-emission 
    (Thiele 2000).
\end{abstract}

\section{The Observations}

    The observational data revealed highly variable absorption structures 
    which could be used to determine the velocity and acceleration structure 
    of the jet. There exists a stable component with a radial velocity 
    of \mbox{$-1100 \pm 200$ km s$^{-1}$} which represents the jet outflow 
    at a larger distance. In addition, strong high velocity components with 
    radial velocities up to \mbox{$-2500$ km s$^{-1}$} appear repeatedly on 
    time scales of a few days and decay within several days. At the same time,
    the absorption between the absorption trough and the emission line with 
    radial velocities of \mbox{$\approx -900$ to $-400$ km s$^{-1}$} 
    disappears which is an indication for an acceleration process near the jet 
    source (see dynamical spectra in Schmid et al. 2001). 

\section{The Models}

    Due to the enormous computational costs, only 2D-models with the 
    assumption of axisymmetry (2.5D) have been calculated yet. Therefore the 
    companion became physically a Red Giant ``Ring''. First, we started a 
    purely hydrodynamical model without cooling. The parameters of this pulsed 
    jet model are listed in Table 1. The influence of the stellar wind to the 
    surrounding medium was simulated with an 1/$r^2$ law. The 
    gravitational potential of the Red Giant and the Jet source was taken into 
    account. The jet pulses were simulated by an enhancement of the jet 
    velocity in the nozzle to \mbox{$2000$ km s$^{-1}$} occurring every seventh 
    day and lasting one day. Thereby also the jet density was varied to 0.25 or
    2 times the normal density.

\begin{table}
\caption{Parameters of the models}
\begin{tabular}{lll}
\tableline
& parameter & value \\
\tableline
jet & number density & $5 \cdot 10^{6}$ cm$^{-3}$ \\
& temperature & 10000 K \\
& velocity & $10^3$ km s$^{-1}$ \\
jet pulse & number density (models I \& II) & $1.25 \cdot 10^{6}$ cm$^{-3}$ \\
& number density (model III) & $1.0 \cdot 10^{7}$ cm$^{-3}$ \\
& velocity (models I - III)& $2 \cdot 10^3$ km s$^{-1}$ \\
stellar wind & $\dot M_{\textrm{red giant}}$ & $1.0 \cdot 10^{-6}$ 
$\textrm{M}_{\odot}$ yr$^{-1}$ \\
& velocity & 10 km s$^{-1}$ \\
system parameters & jet-radius & 1 AU \\
& radius of the Red Giant & 1 AU \\
& binary separation & 4 AU \\
simulation parameters & integration domain & $50 \times 30$ AU \\
& numerical resolution & 20 grid cells/AU \\
\tableline
\tableline
\end{tabular}
\end{table}

    In a second simulation, a cooling mechanism was implemented. According to 
    Sutherland \& Dopita (1993), three species (electrons, protons and 
    hydrogen atoms) were considered and the heavier elements were simulated 
    with a general cooling curve. In a third -- again purely hydrodynamical -- 
    simulation, the number density of the jet pulse was increased to eight 
    times the value of the first two models.

\section{Results}

    While the complete hydrodynamical simulation (450 days) took two weeks on 
    a workstation, after nine weeks on a NEC SX5 vector computer 
    only 74 days of the simulation with cooling were calculated. These 
    computational expenses are caused by the unique parameter choice, 
    especially the high densities, resulting in a small cooling time step. 
    These are the first simulations in this parameter region including cooling.

    Comparing the logarithm of the total density for the same time step, 
    one realizes the effects of cooling: due to smaller pressures in the jet 
    the radial extension of the jet, the cross section and the resistance 
    exerted by the external medium are lessened and therefore the cooled jet 
    propagates faster (Fig. 1).

\begin{figure}
  \plotone{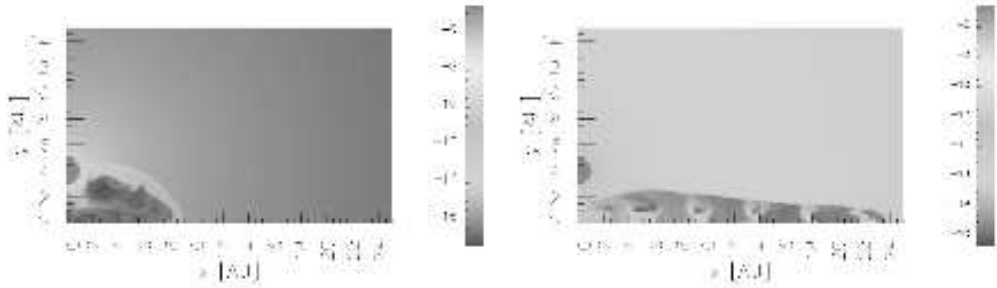}
  \caption{Logarithm of density after 74 days; without (left) and with 
    cooling (right)}
\end{figure}

    Using the density and velocity data, theoretical line profiles can be 
    calculated, assuming the absorption 
    \mbox{$I = I_{0} \, \textrm{e}^{-\tau}$} for each grid cell with

\begin{equation}
\tau_{\lambda} = \frac{\pi\,e^2}{m_{e}\,c}\,\lambda\,N_{j}\,f_{jk} .
\end{equation}

    With the increased number density of the jet pulse, the third simulation 
    is able to reproduce the measured absorption of the fast pulse components 
    (Fig. 2).

\begin{figure}
  \plottwo{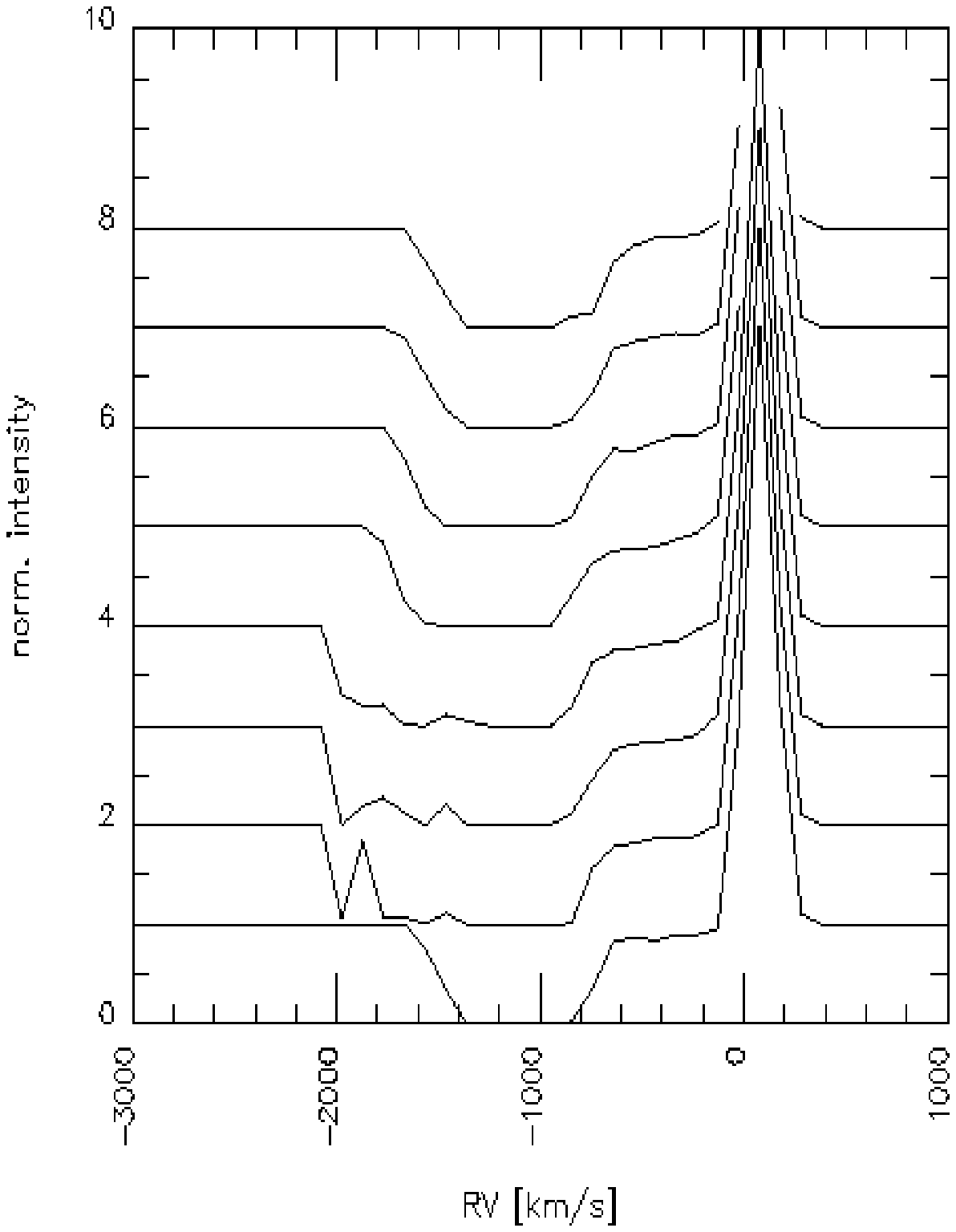}{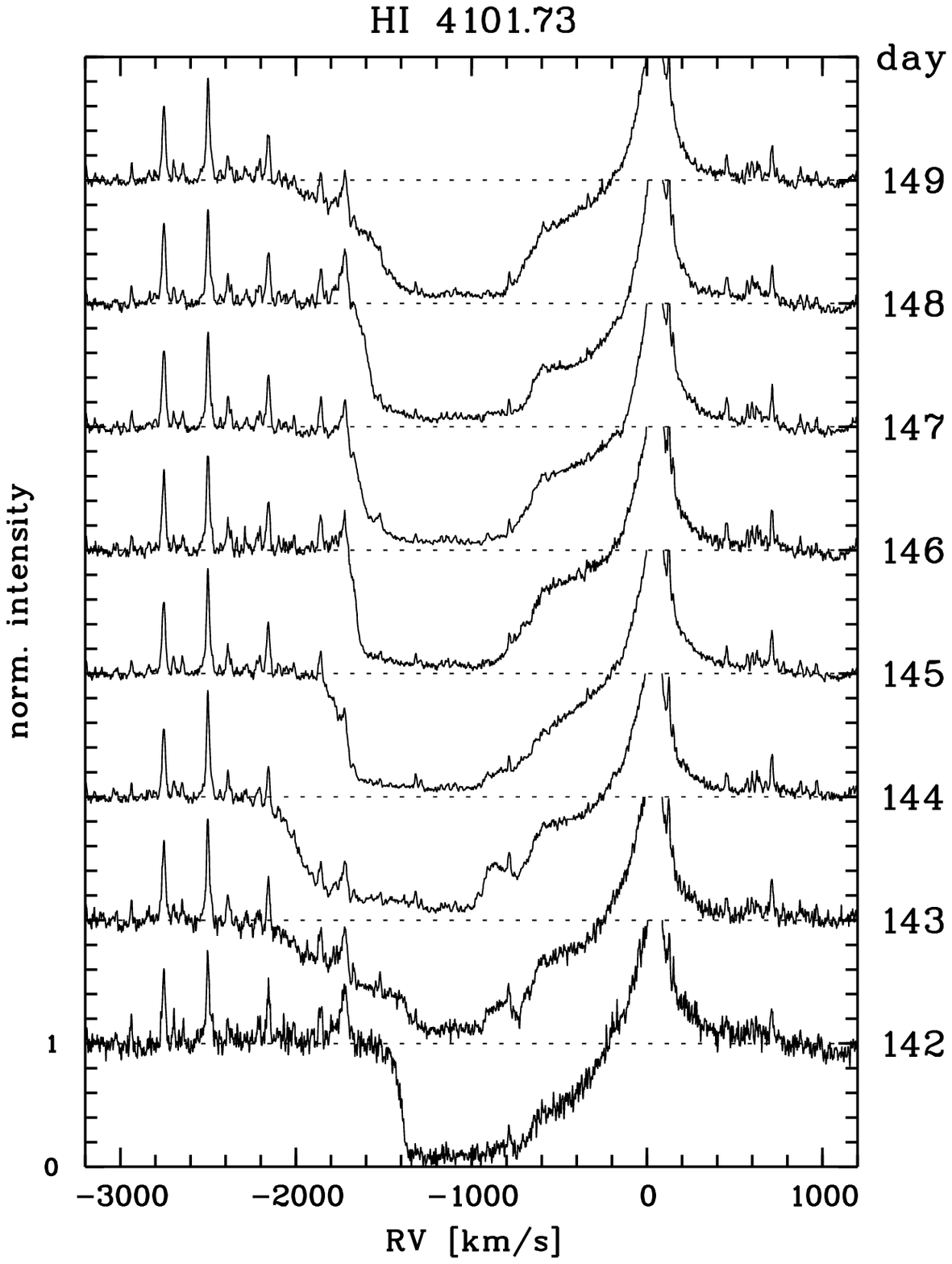}
  \caption{Jet absorption line profiles for a series of eight consecutive 
    days covering a high velocity pulse; theoretical (left) and observed 
    (right)}
\end{figure}

\section{Discussion}

    In the purely hydrodynamical model the high velocity component decays much 
    faster than in the observations -- on time scales of one day. Taking into 
    account the effects of cooling, it should decay slower which would be in 
    agreement with the observational data. Due to the increasing propagation 
    velocity of the jet and the decreasing resistance, the faster component 
    should be able to exist longer. As this simulation is just at its 
    beginning -- and the jet has not yet reached its quasi-stationary phase, 
    the results are still preliminary. But one first result could be the 
    importance of cooling effects for modeling the jet correctly. 

    A parameter study varying further the density and the injection velocity 
    during the outbursts should bring deeper insights into the real mass loss 
    rate and jet energy in this system.


\begin{references}
\reference Sutherland, R.\ S., \& Dopita, M.\ A. 1993, \apjs, 88, 253

\reference Ziegler, U., \&  Yorke, H. 1997, Comp. Phys. Comm., 101, 54

\reference Thiele, M. 2000, PhD Thesis, University of Heidelberg

\reference Schmid, H.\ M., Kaufer, A., Camenzind, M., Rivinius, Th., Stahl, O.,
Szeifert, T., Tubbesing, S., Wolf, B. 2001, \aap, 377, 206

\end{references}
\end{document}